\documentstyle[12pt]{article}
\newcommand{\be}{\begin{equation}}
\newcommand{\ee}{\end{equation}}
\newcommand{\ba}{\begin{eqnarray*}}
\newcommand{\ea}{\end{eqnarray*}}
\newcommand{\bal}{\begin{eqnarray}}
\newcommand{\eal}{\end{eqnarray}}

\title{New bosonization scheme for spin systems in any dimension}

\author{Alexandra Ilinskaia$^1$ and Kirill Ilinski$^{1,2}$
\\
{\small\it $^{1}$ School of Physics and Space Research, University of
Birmingham,}
\\
{\small\it Birmingham B15 2TT, United Kingdom.} 
\\
{\small\it $^{2}$ Institute of Spectroscopy, Russian Academy of
Sciences,} 
\\
{\small\it Troitsk, Moscow region, 142092, Russian Federation.}
}

\date{  }

\begin{document}
\maketitle
\parindent=0pt

\begin{abstract}
We present a new representation of spin operators in terms of bosonic
creation-annihilation operators. This representation allows us to
formulate
a new field-theoretical description of spin systems which is free of any
constraints. The corresponding  
functional integral representations for thermodynamic quantities are
given
and the application to investigations of Long Range Order in the 
system is discussed. 
\end{abstract}

Key words: spin operators,  constraints, bosonization

PACS numbers: 05.30.-d; 75.10.Lp

\section{Introduction}
There are two sources of motivation to search for bosonic
representations of
spin systems and systems of truncated oscillators: the first is
technical
while the second is of principle. Indeed,
a general problem of any perturbative investigation of spin systems 
is the complicated diagram technique which originate from spin-spin 
commutation relations. On the other hand, for systems with the
Hamiltonian 
formulated in terms of the bosonic or fermionic creation -annihilation 
operators, 
the diagram technique is standard and straightforward. That is why 
we need a bosonic representation for the spin operators
to cast the complicated technique into the common form and use the 
field-theoretical machinery. 
Another set of problems where the bosonic treatment is vital is
when looking for Long Range Order (LRO) in the systems.
It is well-known that LRO is reflected in apearance of anomal averages. 
It is always very tempting to reformulate the 
problem in such a way that the anomalous averages become amplitudes of a 
Bose-condensate of some auxiliary bosons. This was a guidline, for
example, 
in Ref.~\cite{AT} where constraint-free representations was found for
Paulions to predict the Bose-condensation of Frenkel excitons. 
In this paper we go along a similar line and develop a constraint-free 
description 
for arbitrary spin system.
To this end we make use the approach developed for truncated oscillators 
in~\cite{II}.

We have to note that there are several transformations that expresse the
spin 
operators in terms of the bosonic or fermionic
ones~\cite{Tsvelik,Fradkin}. 
However, all of them require either the restriction of the bosonic
Hilbert 
space which leads to constraints for the bosonic system or restrict the 
study to 1D systems. The constraints do not cause
any problem unless the systems are treated exactly.
Since it is very difficult to get exact results for spin systems, some
type of approximations should be used. The most popular approximation
scheme 
is based on the mean field description. At this point drawbacks of the 
constrained description emerge. Indeed, the mean field approximation
does not 
treat local (on-site) constraints in a proper way. It means that instead
of 
many local constraints only one global constraint appears. All together
it 
leads to the problems of the account of unphysical
local fluctuations. This effectively returns us to the local constraints
and
explains the importance of the constraint-free formulation of the
mapping 
from spin systems to bosonic ones.

Similar to the approach of the sigma-model with Wess-Zumino 
term~\cite{Fradkin,W} 
we treat the constraint on the number of particles on each site exactly. 
To do this we use the mapping of the orthogonal 
sum of identical copies of the lattice spin space of states to the 
bosonic space of states. In this mapping spin operators are 
represented in the form of a  power series of the bosonic creation and 
annihilation operators. This  compels us to deal with 
infinite series of different vertices in the diagram technique. 
The choice of 
relevant contributions in such series should be dictated as usually by 
features of the concrete problem. 

\section{Mapping of spins to bosons without constraint}

In this section we will describe the mapping from the system of lattice 
spins to the auxiliary bosonic system. The goal is to escape the 
introduction of a constraint. To do this we will embed an infinite
number of 
copies of the 
finite dimensional space of states in the bosonic space of states and
then 
proceed with
the consideration of this new (auxiliary) bosonic space.

To explain this in detail, let us first of all consider one
degree of freedom (i.e. a single site).  Spin operators obey the
following
commutation relations (for spin m/2):
$$
S^- S^+ - S^+ S^- = 2 S^z\ , \quad (S^+)^{+} = S^- \ , \quad 
(S^+)^{m+1} = (S^-)^{m+1} = 0 \ , 
$$
\be
S^+ = S^x + i S^y \ , \qquad
S^- = S^x - i S^y \ .
\label{alg}
\ee
Operators $S^{+}$, $S^-$ and $S^z$ have the following matrix form 
in the $m+1$-dimensional 
Hilbert space of states ${\cal H}_B$: 
\ba
S^{+} & = & \sqrt{2}
\left(
\begin{array}{cccccc}     
0 & 0 & 0 &...  & 0 & 0 \\
\sqrt{m}& 0 & 0 & ... & 0 & 0 \\
0 & \sqrt{2 (m-1)} & 0 & ... & 0 & 0 \\
0 & 0 & \sqrt{3 (m-2)} & ... & 0 & 0 \\
\vdots & \vdots & \vdots & \vdots & \vdots & \vdots\\
0 & 0 & 0 &... &  \sqrt{m} & 0 
\end{array}
\right) \ ,\\ [0.5cm] 
S^- & = & (S^{+})^{Tr} \ , \qquad  
S^{z}  = diag \left( -\frac{m}{2} + k, \ k=0,\ldots, m \right)
\ea
with basis $\{|0\rangle,|1\rangle,...,|m\rangle\}$ and the obvious 
notations.

Now we introduce 
the infinite orthogonal sum ${\cal H}_b=\oplus \sum_{n=0}^{\infty} 
{\cal H}_{S,n}$ of 
such $m$- dimensional Hilbert spaces ${\cal H}_{S,n}$ with basis
$\{\{|0\rangle,|1\rangle, \ldots,|m-1\rangle \}, \ldots,
\{|nm+1\rangle,|nm+2\rangle, \ldots,|nm+m\rangle\}, \ldots\}$. 
The extensions of 
operators $S^{+}$, $S^-$ and $S^z$ in this space have the 
form: 
\ba
\hat{S}^{+} & = & diag(S^{+},S^{+},...)\ ,  \qquad  
\hat{S^-}=diag(S^-,S^-,...)
\ , \\
[0.5cm] 
\hat{S}^{z} & = & diag(S^{z},S^{z},...) \ .
\ea

It follows that all thermodynamic quantities calculated 
with  operators $\hat{S^{+}}$, $\hat{S^-}$, $\hat{S^z}$ are exactly 
the same as 
those calculated with the original operators $S^{+}$, $S^-$, $S^z$.
Indeed, 
for example,
$$
\langle \hat{S}^{+}\hat{S^-}\rangle\equiv 
\frac{Sp(\hat{S}^{+}\hat{S^-}e^{-\beta 
(E-\mu)\hat{S}^{+}\hat{S^-}})}{Sp(e^{-\beta(E-\mu)\hat{S}^{+}\hat{S^-}})} 
$$
coincides with the same expressions but without hats due to the block 
structure
of our operators (we should add that the partition functions differ by 
an infinite numerical constant which does not affect observable physical 
quantities). The conclusion is still valid if we 
start with a lattice of spins  and then introduce hats for 
the operators.

\section{Bosonic representation for spin operators}

Let us now derive the relations for matrix elements of operators
$\hat{S}^+$,
$\hat{S}^-$ and $\hat{S}^z$. To do this we will 
follow the method proposed by Chernyak in  Ref.~\cite{Ch} for Paulions.
The main point of the method is to use the projection operator on the
vacuum 
state
of the auxiliary boson system, i.e on the vector $|0\rangle $. 
This projection operator ${\cal P}$ has the following expression in 
terms of the bosonic creation and annihilation operators:
$$
{\cal P} = \sum_{l=0}^{\infty}\frac{(-1)^{l}}{l!}
(b^{+})^{l}b^{l} \equiv :\exp(-b^{+}b): \ .
$$
We now can use  this representation to construct the operators
$\hat{S}^{+}$, 
$\hat{S}^-$ and $\hat{S}^z$ which obey algebra~(\ref{alg}).
Indeed, it is easy to check from the matrix form that the following 
relations hold: 
\bal
\hat{S}^{+} & = & \sum_{n=0}^{\infty} \sum_{k=0}^{m-1} (b^{+})^{mn+k+1} 
{\cal P} 
b^{mn+k} \frac{\sqrt{(k+1)(m-k)}}{(mn+k)!\sqrt{mn+k+1}}\ ,
\quad \hat{S}^- = (\hat{S}^+)^+ \nonumber\\
\label{Chernyak}
\hat{S}^{z} & = & \sum_{n=0}^{\infty} \sum_{k=0}^{m} (b^{+})^{mn+k}
{\cal P} 
b^{mn+k} \frac{(-m/2 +k) }{(mn+k)!} \ .  
\eal
It is obvious that these relations satisfy the  
algebra~(\ref{alg}). For the particular case $m=2$ our
formulae reduce to the formulae originally obtained by 
Chernyak~\cite{Ch} for the case of paulionic operators.

\section{Thermodynamics in functional integral representation}

The formulae considered above can be applied to construct 
the Hamiltonian of the auxiliary bosonic system.
Let us start with the following  Hamiltonian $H_S$ of spins on a
lattice: 
\ba
H_S & = & \sum_{i\neq k} X_{ik} S^{x}_{i} S^x_{k} + 
\sum_{i\neq k} Y_{ik} S^{y}_{i} S^y_{k} + 
\sum_{i\neq k} Z_{ik} S^{z}_{i} S^z_{k} + \\
 & + & \sum_{i} (h_{ix} S^{x}_{i} + h_{iy} S^{y}_{i} + h_{iy} S^{z}_{i})
\ .  
\ea
Using operators $S^+$, $S^-$ one can cast it in the following form:
\ba
H_S & = & \frac{1}{2} \sum_{i\neq k} (X_{ik} + Y_{ik}) S^{+}_{i} S^-_{k}
+ 
\frac{1}{4} \sum_{i\neq k} 
\left\{(X_{ik} - Y_{ik}) S^{+}_{i} S_{k}^{+} + h.c. \right\} + \\
& + & \sum_{i\neq k} Z_{ik} S^{z}_{i} S^{z}_{k} +
\frac{1}{2} \sum_{i} 
\left\{(h_{ix} - i h_{iy}) S^{+}_{i} + h.c. \right\} +
\sum_{i} h_{iz} S^{z}_{i}\ .   
\ea
The corresponding Hamiltonian of the auxiliary bosons based 
on the relations~(\ref{Chernyak}) has the form:
\ba
H & = &
\frac{1}{2} \sum_{i\neq k} (X_{ik} + Y_{ik}) b^{+}_{i} S_{ik} b_{k} + 
\frac{1}{4} \sum_{i\neq k} 
\left\{ (X_{ik} - Y_{ik}) b^{+}_{i} b_{k}^{+} S_{ik} + h.c. \right\} +
\\ 
 & + & \sum_{i\neq k} Z_{ik}
\sum_{l,m=0}^{\infty}a(l)a(m)(b^{+}_{i})^{l} 
(b^{+}_{k})^{m} b_{i}^{l} b_{k}^{m} + \\
 & + & \frac{1}{2} \sum_{i}  \left\{  (h_{ix} - i h_{iy}) 
 \sum_{l=0}^{\infty} A(l) (b^{+}_{i})^{l+1} b_i^l + h.c. \right\} +
\sum_{i} h_{iz} \sum_{l=0}^{\infty} a(l) 
(b^{+}_{i})^{l} b_{i}^{l} \ .
\ea
Here the following notations have been introduced:
\ba
S_{ik} & = & \sum_{l,m=0}^{\infty} A(l) A(m)
(b^{+}_{i})^{l} (b^{+}_{k})^{m} b_{i}^{l} b_{k}^{m} \ , \\[0.5cm]
A(l) & \equiv & \sqrt{2}\sum_{k=0}^{min(m-1,l)}
 \sum_{n=0}^{\left[ \frac{l-k}{m} \right]} 
\frac{(-1)^{l-mn-k}}{(l-mn-k)!} 
\frac{\sqrt{(k+1)(m-k)}}{(mn+k)! \sqrt{mn+k+1}} \ ,  \\[0.5cm]
a(l) & \equiv & \sum_{k=0}^{min(m,l)} \sum_{n=0}^{\left[ \frac{l-k}{m} 
\right]} 
\frac{(-1)^{l-mn-k}}{(l-mn-k)!} 
\frac{(-m/2+1)}{(mn+k)!} \ . 
\ea

Using the standard procedure, we can write down the functional integral 
representation of the partition function and correlators of the
auxiliary 
bosonic system and the original system of truncated oscillators. For 
example, according to the definition and  the formula (\ref{Chernyak}), 
the following relations arise:
\ba
Z &\equiv& Sp(e^{-\beta{H}}) = 
\int Db^{+}(t)Db(t) e^{S} \ , \\[0.5cm]
\langle {S}^{+}_{i}{S}^{-}_{k} \rangle &=& \int Db^{+}(t)Db(t) 
\sum_{l,m=0}^{\infty} A(l)A(m)(b^{+}_{i}(t))^{l+1} 
(b^{+}_{k}(t))^{m}b_{i}^{l}(t)) b^{m+1}_{k}(t) e^{S}/ Z \ ,
\ea
where the action $S$ is defined by the form of the Hamiltonian $H$:
\ba
S & = & \int_{0}^{\beta} d t \Biggl( \sum_{i} \frac{\partial 
b^{+}_{i}(t)}{\partial t} b_{i}(t) - 
\frac{1}{2} \sum_{i\neq k} (X_{ik} + Y_{ik}) b^{+}_{i}(t) S_{ik}(t) 
b_{k}(t) \\
 & - & \frac{1}{4} \sum_{i\neq k} 
\left\{ (X_{ik} - Y_{ik}) b^{+}_{i}(t) b_{k}^{+}(t) S_{ij}(t) + h.c. 
\right\}\\ 
 & - & \sum_{i\neq k} Z_{ik} 
 \sum_{l,m=0}^{\infty}a(l)a(m)(b^{+}_{i}(t))^{l} 
(b^{+}_{k}(t))^{m} b_{i}^{l}(t) b_{k}^{m}(t)  \\
 & - & \frac{1}{2} \sum_{i}  \left\{  (h_{ix} - i h_{iy}) 
 \sum_{l=0}^{\infty} A(l) (b^{+}_{i}(t))^{l+1} b_i^l(t) + 
 h.c. \right\} \\
 & - &
\sum_{i} h_{iz} \sum_{l=0}^{\infty} a(l)(b^{+}_{i}(t))^{l} b_{i}^{l}(t)
\Biggr)\ .
\ea

All other correlators can be obtained in the same manner and give us 
the bosonic functional integral 
representation which is free of  constraints and limiting procedures
The functional integral form then allows 
the simplest approach to the derivation of diagram technique rules which
are 
standard ones for the problems in question. It is tempting to 
note that this technique is much less complicated and much more 
straightforward
than the spin operator technique and is very natural for the
consideration 
of problems concerning Bose-condensation (Long Range Order) in the
system 
just using the standard Bogoliubov's approach to the subject.

\section*{Acknowledgments}

This work was supported by the UK EPSRC Grant GR/L29156 and
the Royal Society/NATO Postdoctoral Fellowship
Award.

\end{document}